\documentclass[10pt,onecolumn,floatfix]{revtex4-1}
\usepackage{hyperref}
\usepackage{graphicx}
\usepackage{epstopdf}
\usepackage{amssymb}
\usepackage{mathrsfs}
\usepackage{amsmath}
\usepackage{color}
\usepackage{latexsym}
\usepackage{amsfonts}
\usepackage{subfigure}
\usepackage{wasysym}
\usepackage{dcolumn}
\usepackage{bm}
\usepackage{subfigure}
\usepackage{natbib}
\usepackage{tabularx}
\usepackage{booktabs}
\usepackage{rotating}
\begin{document}

\title{Universal criterion for designability of heteropolymers}
\author{Chiara Cardelli} 
\affiliation{Faculty of Physics, University of Vienna, Boltzmanngasse 5, A-1090 Vienna, Austria}
\author{Valentino Bianco} 
\affiliation{Faculty of Physics, University of Vienna, Boltzmanngasse 5, A-1090 Vienna, Austria}
\author{Lorenzo Rovigatti} 
\affiliation{Faculty of Physics, University of Vienna, Boltzmanngasse 5, A-1090 Vienna, Austria}
\author{Francesca Nerattini} 
\author{Luca Tubiana} 
\author{Christoph Dellago}
\affiliation{Faculty of Physics, University of Vienna, Boltzmanngasse 5, A-1090 Vienna, Austria}
\author{Ivan Coluzza}
\affiliation{Faculty of Physics, University of Vienna, Boltzmanngasse 5, A-1090 Vienna, Austria}



\begin{abstract}
\noindent Proteins are an example of heteropolymers able to self-assemble in specific target structures. The self-assembly of designed artificial heteropolymers is still, to the best of our knowledge, not possible with control over the single chain self-assembling properties comparable to what natural proteins can achieve. What artificial heteropolymers lacks compared to bio-heteropolymers that grants the latter such a versatility? Is the geometry of the protein skeleton the only a particular choice to be designable?
Here we introduce a general criteria to discriminate which polymer backbones can be designed to adopt a predetermined structure. 
With our approach we can explore different polymer backbones and different amino acids alphabets. 
By comparing the radial distribution functions of designable and not-designable scenarios we identify as designability criteria the presence of a particular peak in the radial distribution function that dominates over the random packing of the heteropolymer. We show that the peak is a universal feature of all designable heteropolymers, as it is dominating also the radial distribution function of natural proteins. Our finding can help in understanding the key features that make proteins a highly designable system.
The criteria that we present can be applied to engineer new types of self-assembling modular polymers that will open new applications for polymer based material science.
\end{abstract}
\maketitle
\noindent  The control of the self-assembly is key in the generation of smart materials, with applications ranging from energetics~\cite{Umena2011}, to photonic crystals~\cite{furumi::jmatchemc::1::2013} and biomimetic scaffolding~\cite{yang::advmat::20::2008}.  Heteropolymers are an important example of self-assembling systems. The technology for their synthesis and manipulation is already advanced~\cite{Lodge2003,Capone2010}, but it is still not possible to design them with control over the single chain structure comparable to natural bio-polymers, such as DNA~\cite{Rothemund2006} and proteins~\cite{Chino2015,Huang2014a}. Here we introduce a sufficiency criterion to engineer heteropolymers with equilibrium structures directly encoded into the building blocks, the so-called \textit{bottom-up} approach to self-assembly. The criterion is based on the appearance of a particular peak in the radial distribution function $g(R)$ that dominates over the random packing of the heteropolymer. Moreover, we show that the peak is a universal feature of designable heteropolymers, as it is dominating also the $g(R)$ of natural proteins.
In fact, natural systems such as DNA~\cite{Rothemund2006}, RNA~\cite{Conde2015} and, in particular, proteins~\cite{Chino2015,Huang2014a}, are the most versatile example of heteropolymers with self-assembling properties controlled via a variable pattern (sequence) of a fixed alphabet of chemically different building blocks (monomers). The specific sequence drives a heteropolymer to uniquely collapse (fold) into a target conformation, with a precise control of the structure. In the case of proteins, different sequences of the same alphabet of 20 amino acids lead to the huge number of proteins expressed in nature, making the same alphabet of building blocks extremely versatile. \\

According to mean field theories (MFT)~\cite{Gutin1993a,Shakhnovich1993a,Pande2000} it should be possible to construct artificial heteropolymers that, similarly to proteins, drive the collapse of specific sequences into highly arbitrary structures. Unfortunately, with the exception of biopolymers~\cite{Gonen2015,Huang2014a,Coluzza2014}, no other polymer architectures are known to exhibit this behaviour. Recently attempts to overcome this limitation have been done by Khokhlov and collaborators~\cite{Khalatur2007} and more recently by Moreno et al.~\cite{moreno2013advantages}. However, contrary to biopolymers, this methodology does not provide control over the detailed shape of the target structure. In other words, it is extremely difficult to identify patterns made of artificial monomers that drive a single chain towards a very specific structure. Since the identification of such patterns is normally referred to as \textit{design}, in what follows we will define \textit{designability} as the property of a heteropolymer to have at least one heterogeneous pattern that folds into one given target structure~\cite{Pande2000}. 

\begin{figure}[t!]
\includegraphics[width=0.13\textwidth]{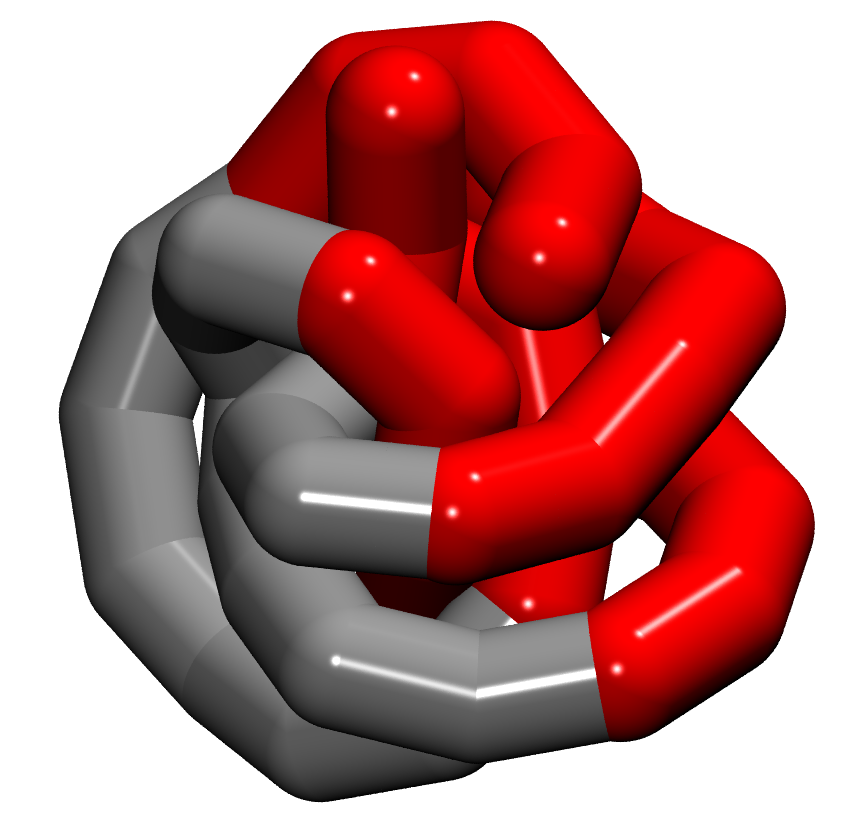} \hspace{0.9cm}
\includegraphics[width=0.13\textwidth]{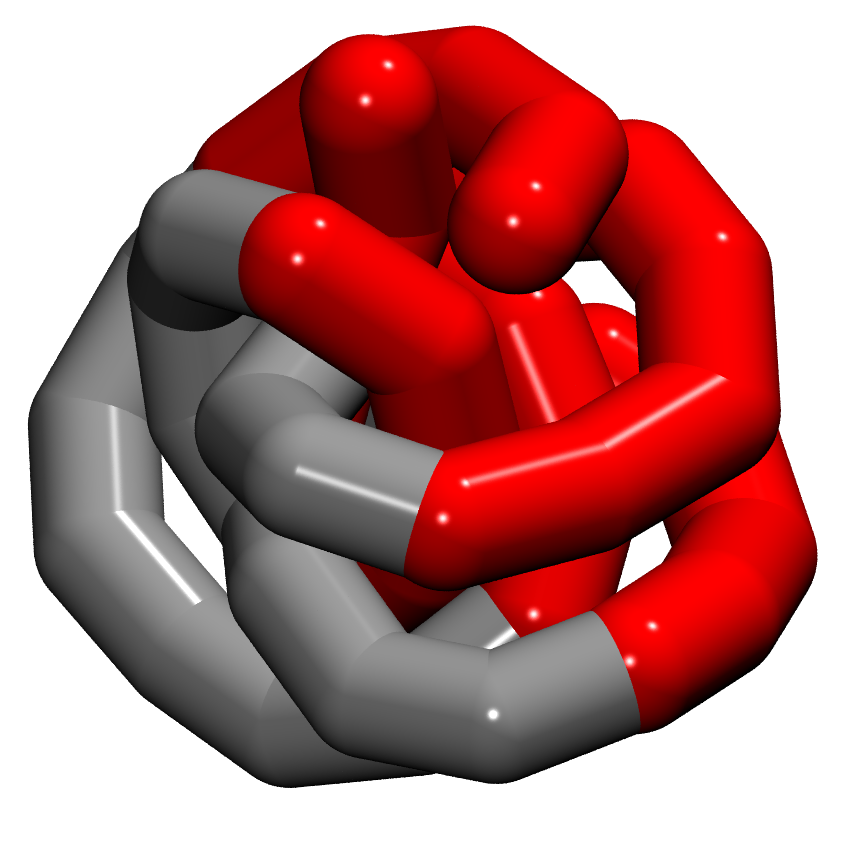} \\
\includegraphics[width=0.13\textwidth]{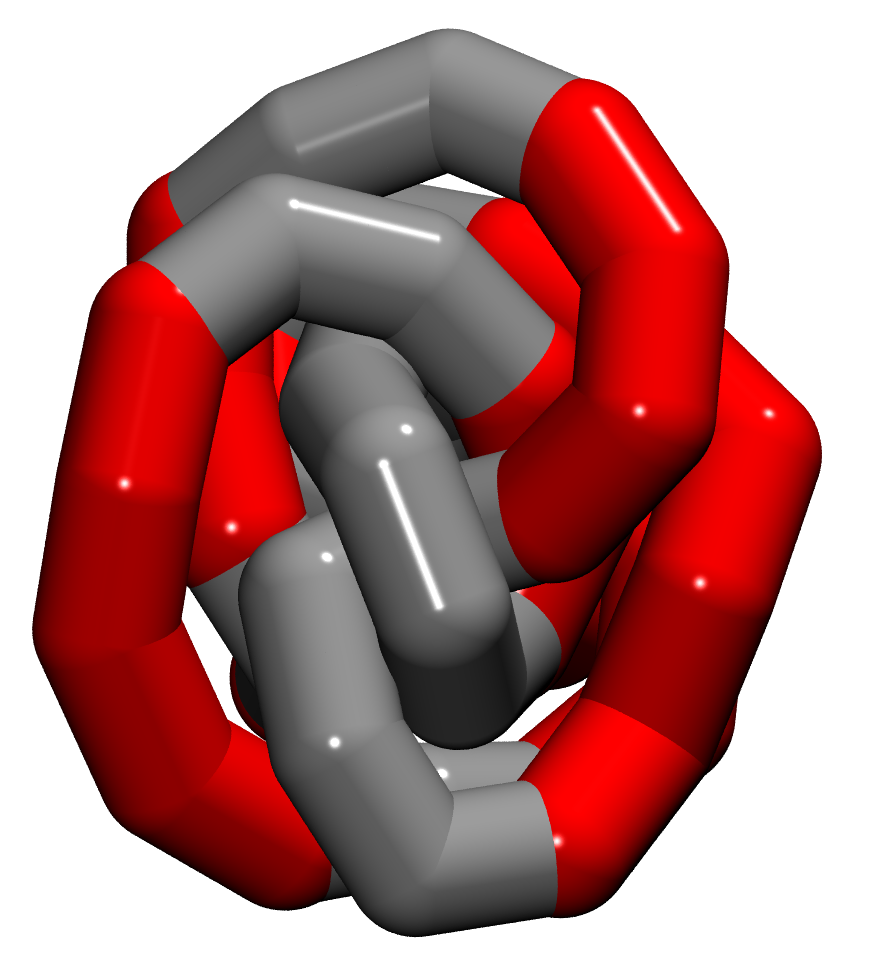} \hspace{0.9cm}
\includegraphics[width=0.13\textwidth]{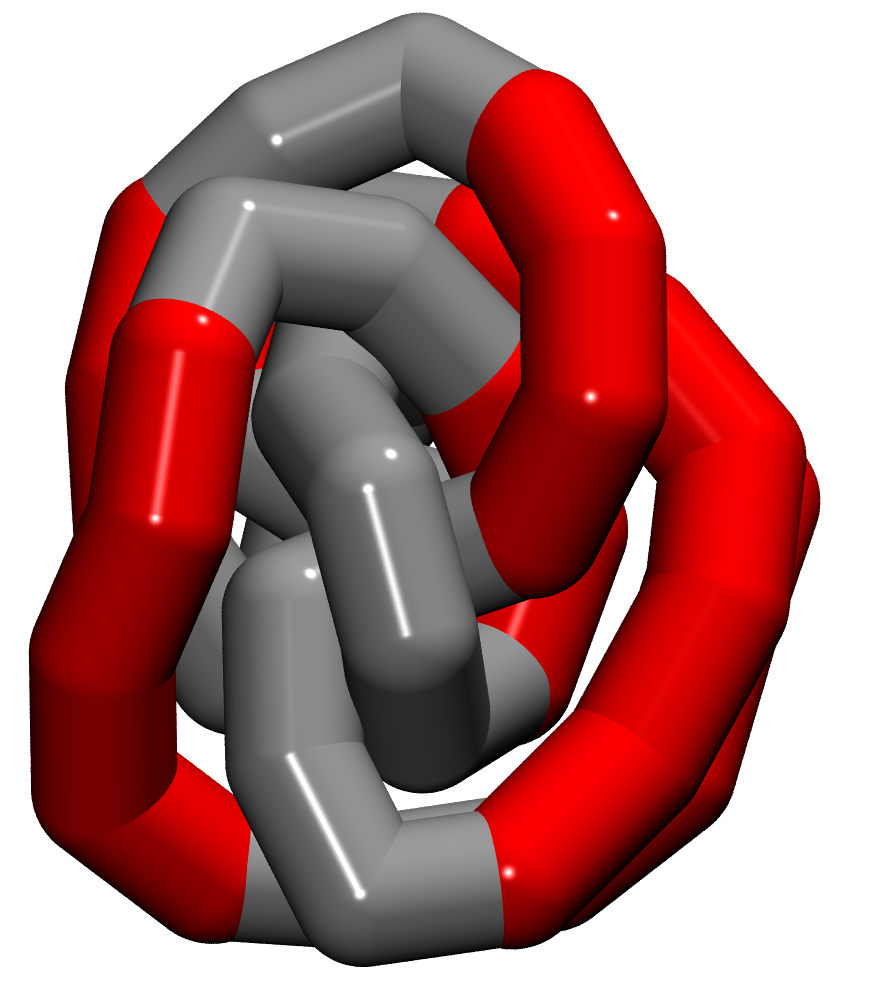} \\
\caption{\label{fig:patterns} Functionalized patchy polymer structures. The structures on the left side are the target structures with the functionalizations we want to achieve, the ones on the right are the final folded structures we obtain. Such functionalized polymers can be used themselves as particles with one (upper structures) or two (lower structures) patches in a hierarchical self-assembling process. The accuracy necessary for the different applications must be of the order of a fraction of the distance between particles. For instance, in the case of natural catalyzers such as the proteins, the atom positions that enhance the target chemical reaction are usually conserved within a fraction of amino acids distances. The red coloured particles identify the target active site that is reproduced with an accuracy of approximately$~\sim0.25$ of the inter-particle distances.}
\end{figure}

Here we show that the addition of the directional interactions on a generalized heteropolymer allows for the realization of compact structures with functionalized regions on their surfaces (see Fig.~\ref{fig:patterns}) with high precision in the folding of approximately$~\sim0.25$ of the inter-particle distances. Depending on the chemical nature of the functionalization, the assembled structures could have catalytic functions or be used themselves as building blocks in a hierarchical self-assembling process. To prove the emergence of the control we employ the ``patchy polymer'' model ~\cite{Coluzza2013,Coluzza2012c} (see Methods) in which directional interactions (patches) are added to the isotropically interacting monomers.

\begin{figure*}[htp]
 \leftskip 4.7 cm 
 {\bf (a)\hspace{4.7 cm}(b)}\\
 \centering
\includegraphics[width=0.4\textwidth]{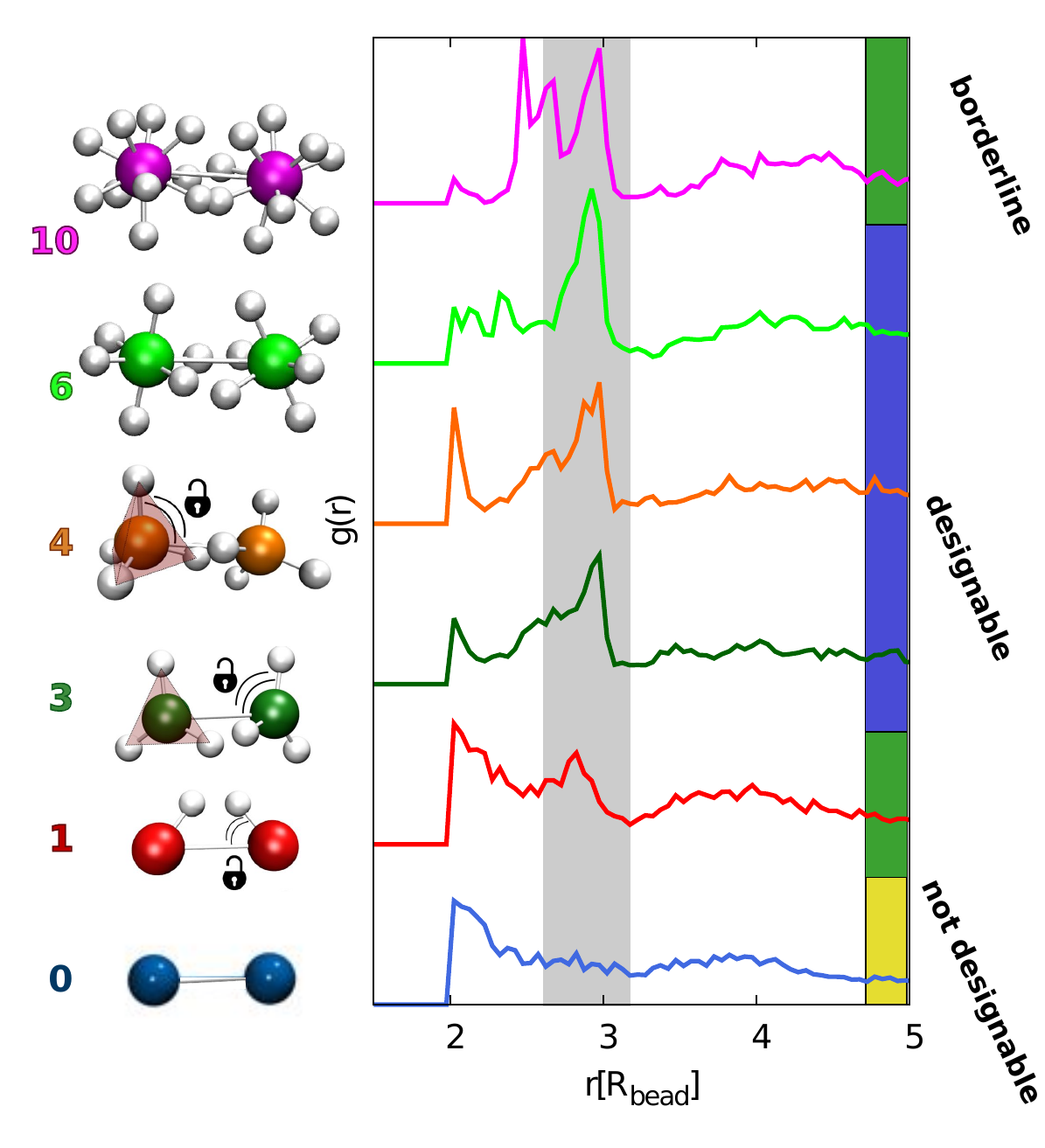}
\hspace{-0.1 cm}
\includegraphics[width=0.3338\textwidth]{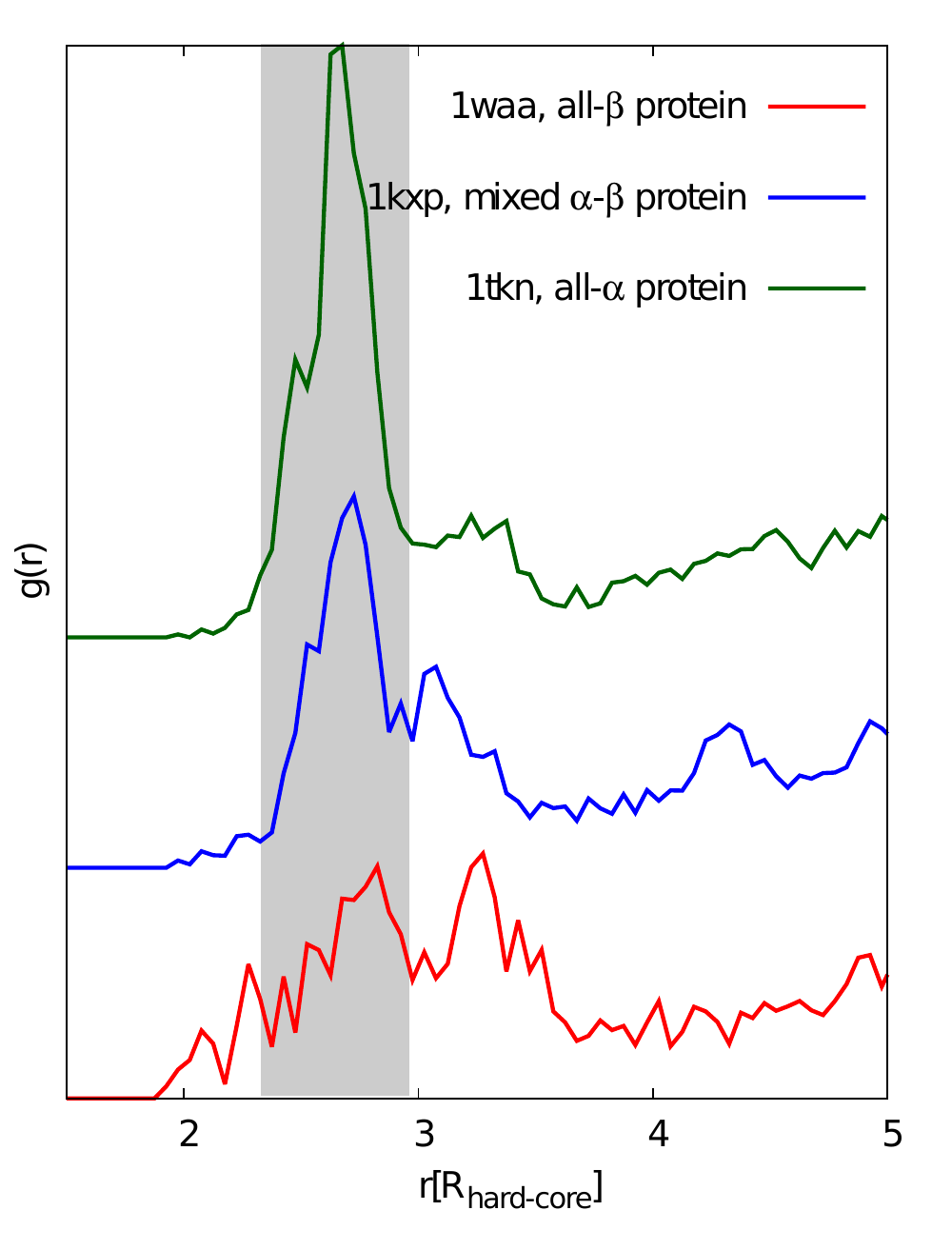}
\vspace{-0.5 cm}
\caption{ a) Radial distribution functions $g(r)$ for patchy polymers. The $g(r)$ is represented as a function of the bead centres distance, for different patches numbers. On the y-axis are shown the schematic representations of the patches structures: small spheres indicate the patches located at the surface of the central bead. The spring between the beads is bonded at the centre of each bead and the patches are free to rotate with respect to it but fixed with respect to each others - we name this model as freely rotating chain. We study this system for $n=0,1,3,4,6,10$ number of patches, placed on an equilateral triangle for the $n=3$ and on tetrahedron for the $n=4$ configurations. For $n>4$ see Supplementary Information. The curves are y-shifted for the sake of visualization. The grey band highlights the range of the characteristic distances of the directional interactions.
All the $g(r)$ have been calculated by neglecting the contribution of first neighbours along the backbone.
b) $g(r)$ of three proteins, representative of the 20 analyzed, taken from the Protein Data Bank (PDB) characterized by only $\alpha$-helices (PDB id. 1waa), only $\beta$-sheets (PDB id. 1tkn) and mixed $\alpha$-helices---$\beta$-sheets (PDB id. 1kxp), each averaged over tens of alternative experimentally proposed structures.
The first peaks dominating the $g(r)$ of each protein correspond to the characteristic distances of the backbone hydrogen bonds (Supplementary Figure~\ref{seek}). Hence, since the compact conformations of protein backbones are sculpted by the geometry of the directional interactions, we expect the protein hydrogen bonds to be an ideal template for highly designable heteropolymers.}
\label{fig:gr}
\end{figure*}

\begin{figure*}[htp]
\centering
\includegraphics[width=1\columnwidth]{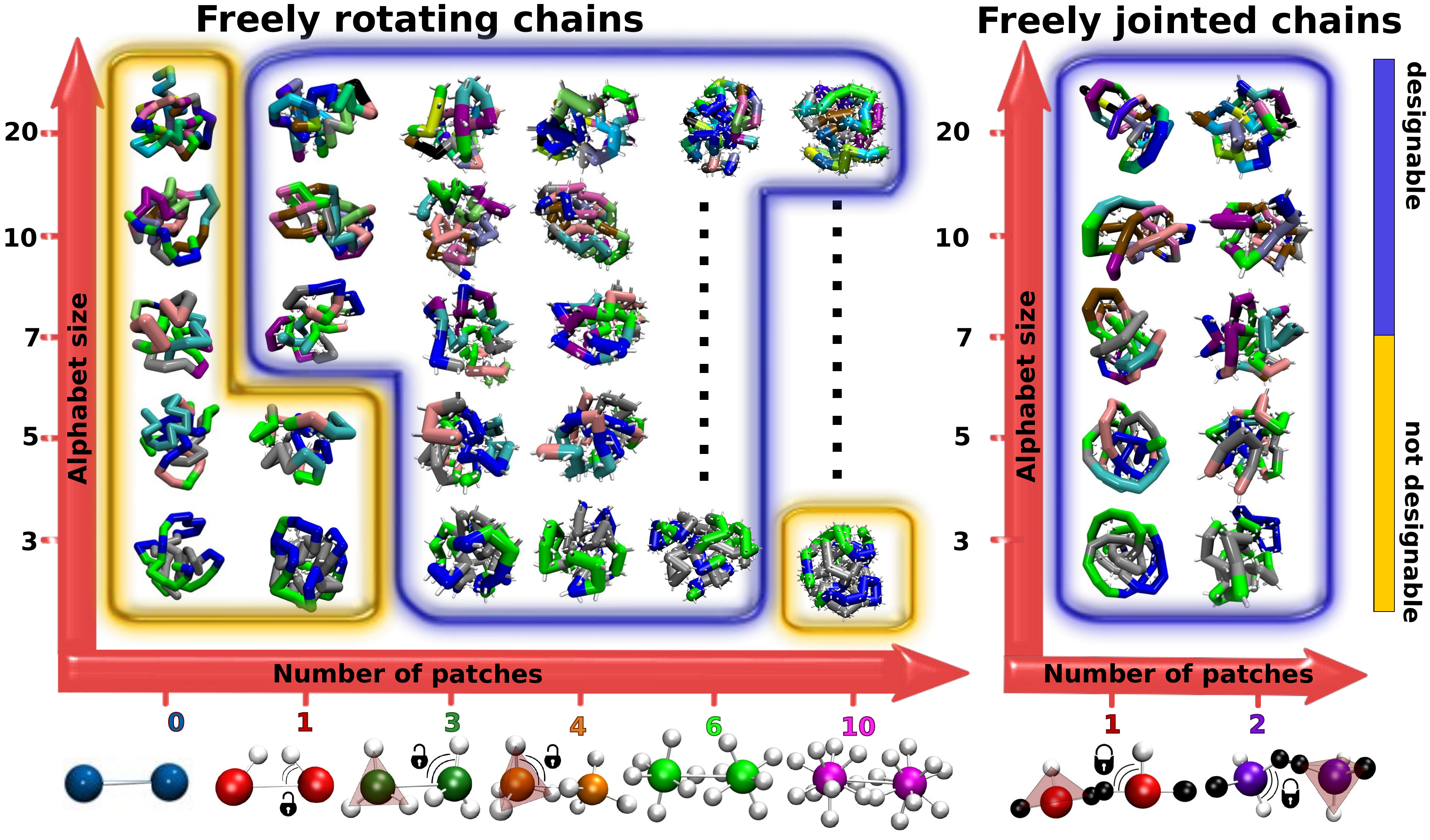} 
\caption{Designability diagram for different patches numbers and alphabet sizes. Within the yellow regions the heteropolymer is not designable, while the blue regions correspond to designable systems. For each case, the chosen target structure is shown. Bottom panel: schematic illustration of patchy polymers. Small spheres indicate the patches located at the surface of the central bead. Left: Freely rotating chains, see Fig.~\ref{fig:gr}. Right: Freely jointed chains: the spring between the beads connects two patches (the black ones) opposite to each others with respect to the centres of the spheres. The number of free patches is thus $n - 2$ (the white ones).}
\label{fig:diagram}
\end{figure*}

However, the question remains open: ``Can we predict \textit{a priori} whether an engineered heteropolymer is designable?''
Here we show that the answer to this question is represented by a distinct peak in the radial distribution function dominating over the random packing of the heteropolymes. A minimal set of directional interactions is an effective way to induce such a peak (Fig.~\ref{fig:gr}), and is sufficient to guarantee designability (Fig.~\ref{fig:diagram}).

Strictly speaking, proving that a specific heteropolymer configuration is not designable would require testing all possible heterogeneous sequences for the presence of a unique collapsed equilibrium structure. Here we rely on the statistical definition of designability given in~\cite{Coluzza2012c}, where the methodology of the Monte Carlo simulations SEEK, DESIGN and FOLD (SDF) was proven to be able to discriminate between designable and not-designable structures. Briefly, the SDF method seeks the most designable target structure (Supplementary Figure~\ref{seek}), designs the sequence that should optimally fold into the target structure and then tests whether it correctly folds or not. If the folding procedure fails, the system is labelled as non-designable, because it was unable to fold into the most favourable target structure (for details see~Methods \ref{sec:methods}).
Here we apply the SDF method to chains made of $50$ monomers, with number of patches ranging from 0 to 10 and an alphabet of isotropic interactions of size 3 to 20 (see~\ref{sec:methods} for the definition of the interactions). In fact, changing the number of patches per monomer makes it possible to move from a simple heteropolymer (0 patches) to a patchy polymer. From now on, all the distances will be in units of the bead radius $R_{bead}$ and the energies in units of $k_BT_{Ref}$, where $T_{Ref}$ is a reference temperature that sets the scale of interactions. 

In Fig.~\ref{fig:diagram} we show the results for the designability for all cases studied. The systems without directional interactions (patches) are not designable for any alphabet. By adding directional interactions, the heteropolymer becomes designable for a wide range of patch numbers and alphabet sizes, both for freely rotating chains and freely jointed chains. The emergence of designability coincides with the appearance of a peak in the pair radial distribution function---highlighted by the grey band in Fig. \ref{fig:gr}a)---located at the distance $r\simeq3$ between the bead centres, at which the patch-patch interaction is most favourable. The presence of such a isolated intermediate peak between the first ($r\simeq2$) and second random close packing neighbours ($r\simeq4$) indicates that the directional interactions are inducing a geometrical frustration in the system. The frustration strongly biases towards a subset among the compact configurations. In fact, in the non-designable configurations without patches the directional interaction peak is not present. In the configurations with one patch, the directional interaction peak is present but is lower than the peak corresponding to the close packing. These configurations are not always designable (borderline), and their designability depends on the alphabet size (Fig.~\ref{fig:diagram}). Increasing the number of patches the directional interaction peak becomes dominating over the random packing peak, and hence the configurations become designable for each alphabet. Increasing further the number of patches up to 10, the system loses directionality as the directional peak splits into more components (Fig.~\ref{fig:gr}a) and this corresponds to a loss in designability (Fig.~\ref{fig:diagram}). It is important to stress that the strength of the directional interaction is such that even the single patch configuration is always maximally bonded, implying that arbitrarily increasing the relative strength of the directional interactions will not suppress the first peak of the $g(r)$ (Supplementary Figure~\ref{seek}). Thus, when the patch-patch peak dominates over the random close packing peak, the system becomes robustly designable for all alphabet sizes, both for freely jointed and freely rotating chains. 

According to the MFT of protein design~\cite{Pande2000}, in order to make a system designable, the alphabet size $q$ must be sufficiently large compared to the conformational entropy per particle $\omega$, or formally: $\ln(q)> \omega$. Hence, the designable--to--not-designable transition, where $\omega\sim\ln(q)$, allows us to estimate $\omega$. Thus, following the diagram in Fig.~\ref{fig:diagram}, we estimate $\omega$ to be $>\ln(20)$ for 0 patches, $\sim\ln(5)$ for one patch in the freely rotating chain, and $<\ln(3)$ for up to 6 patches.  At 10 patches  $\omega$ increases again to a value between $\ln(3)$ and $\ln(20)$. Since $e^{\omega}$ is connected to the number of compact conformations, the latter is reduced by the patches by approximately an order of magnitude. A more precise evaluation of $\omega$ requires the study of intermediate scenarios and is object of current investigation. 

The presence of the directional interaction peak is a fundamental fingerprint of designability. Indeed, we find it to be a general feature also of natural proteins. In Fig.~\ref{fig:gr}b we show the radial distribution function for some characteristic examples out of 20 analysed natural proteins. Here, the conformational space is shaped by the directionality of the hydrogen bonds, which forces the carbon C$_{\alpha}$ of the amino acids to be at the typical distances for the different types of secondary structure. Hence, the peaks highlighted in grey in Fig.~\ref{fig:gr}b are equivalent to the directional interaction peak observed in the patchy polymers (Fig.~\ref{fig:protein_scheme} in the Supplementary Information for more details).

\begin{figure*}[t]
\centering
\includegraphics[width=1\textwidth]{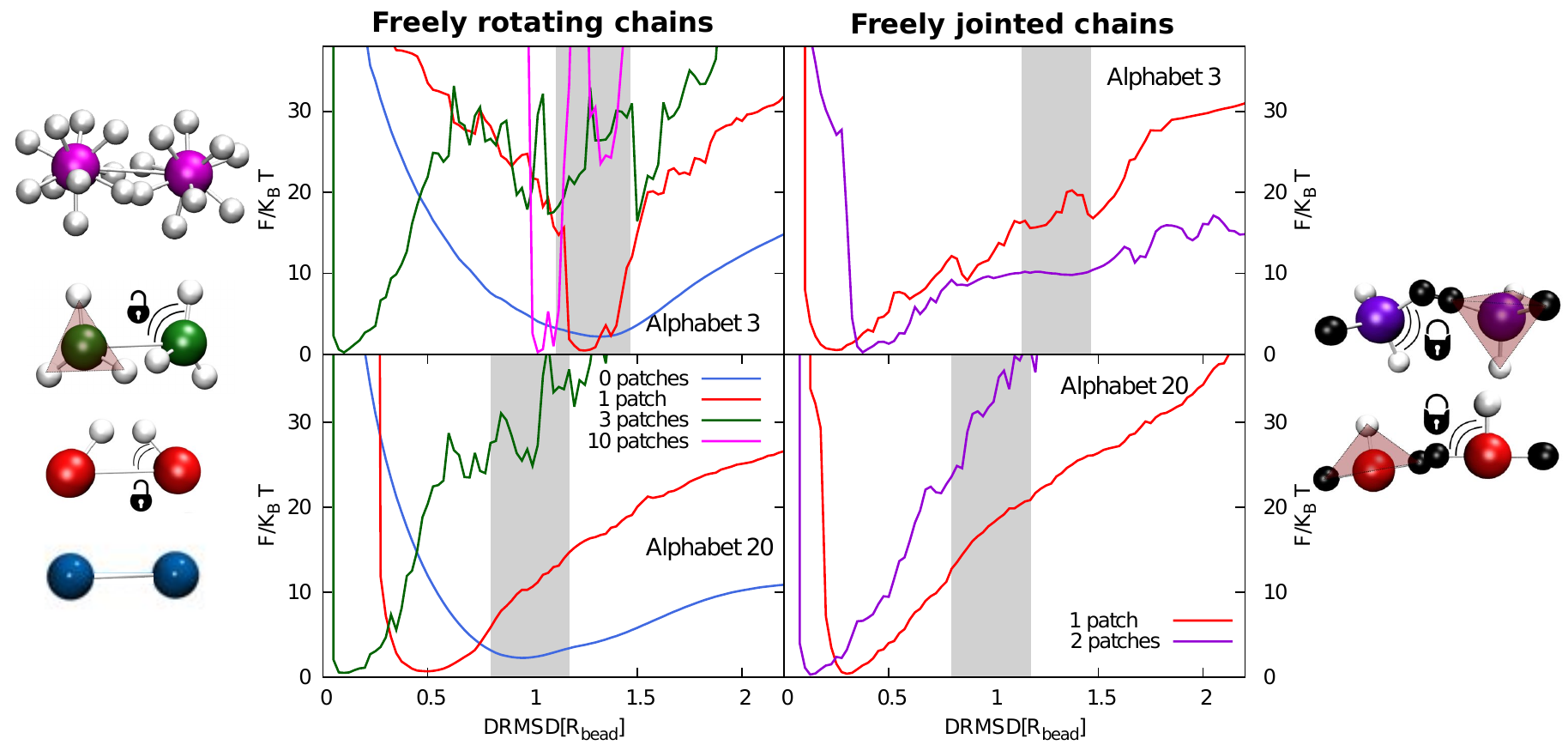}
\caption{ FOLDING free energy landscapes for some significant cases. The free energy is plotted as a function of the distance root mean square displacement ($DRMSD$) for freely rotating chains (left panels) and freely jointed chains (right panels), for different patch numbers and alphabet sizes at temperature $0.4$. $DRMSD=0$ corresponds to the target structure. The molten globule region is highlighted with a grey area, and all global minima with $DRMSD$ values below the grey area are considered correctly folded (for details see~Methods \ref{sec:methods}). The geometry of the freely rotating patches and freely jointed patches are reported on the left and right side of the figure, respectively, with colours corresponding to the free energy curves. The global minimum of the system with 10 patches and alphabet 3 is at the border with the grey area and it has been categorized as not designable.}
\label{fig:results}
\end{figure*}

These results are in agreement with MFT~\cite{Gutin1993a,Shakhnovich1993a,Pande2000}, which predict that the decrease of conformational freedom allows for better sequence design. Here, we introduce the presence of a peak in the $g(R)$ not related to the random packing as an estimate of the reduction of the conformational freedom, in order to have a simple tool to anticipate the designability of different polymer architectures. Thus, we propose the presence of such a peak as a general criterion to engineer designable polymer architectures able to fold into unique target structures, with the same accuracy and versatility of natural proteins. The geometry of the protein skeleton is a particular choice, and our results suggest that, by following the criterion of the directional interaction peak, several others can be found.

Another important result is the characterization of the transition from not-designable to designable configurations (Fig.~\ref{fig:diagram}). We start by noting that the freely rotating chain is sensitive to the choice of the alphabet: a minimum alphabet size of 7 is required to guarantee designability for any configurations of the patches (Fig.~\ref{fig:diagram}).
To characterize the designable--to--not-designable transition, we have performed for each point in the designability diagram the SDF trial. The last step of SDF is the calculation of the free energy difference between structures, grouped together according to the distance root mean square displacement ($DRMSD$) from the target conformation at $DRMSD=0$ (Methods \ref{sec:methods}). The accuracy of the refolding varies considerably for each scenario, and depends on the alphabet size, the number of patches and the local environment of the beads. In Fig.~\ref{fig:results} we show the FOLDING free energy profiles for some significant points in the designability diagram for alphabet sizes 3 and 20 (other alphabets are shown in Fig.~\ref{fig:folding_total_results} in the Supplementary Information). At low temperature ($T=0.4$) all curves show a global free energy minimum located at different $DRMSD$, however they are considered folded only if the global minimum corresponds to the folded structure and not to a molten globule disordered structure (for details see~Methods \ref{sec:methods}). What is striking in the figure is the high refolding accuracy of the freely jointed chain model, with smooth folding profiles and global minima very close to the target structure. We ascribe such a high accuracy to the stronger constraint experienced by monomers in a freely jointed chain compared to the freely rotating model, in agreement with MFT results~\cite{Gutin1993a,Shakhnovich1993a,Pande2000}. In fact, the patches in the freely jointed chain have less rotational freedom, thus the conformational entropy $\omega$ is decreased further. Finally, we note that such high accuracy is particularly interesting for single-patch chains that, according to the designability criterion based on the $g(r)$, is a borderline case (Fig.~\ref{fig:gr}a).

In conclusion, we extend the concept of \textit{designability} to heteropolymers, bridging the gap between naturally foldable biopolymers and artificial polymers. We demonstrate that there is a minimal set of ingredients that makes it possible to exert a precise control over the conformation of the folded structure. 
Indeed, we show that, starting from a traditional heteropolymer model (0 patches), we can attain designability by introducing a few directional interactions. The patches make the system designable even with alphabet sizes comparable to the case of block-copolymers (3 and 5). It is important to stress that the technology for the synthesis and manipulation of artificial heteropolymers is at an advanced stage, as demonstrated by the numerous materials based on block copolymers~\cite{Lodge2003,ISI:000253221300007,ISI:000254135200022,ISI:000259589300026}. We demonstrate that the directional interaction peak in the $g(R)$ is a universal fingerprint for designability and thus can be used as a criterion to engineer new heteropolymers. This criterion will allow artificial heteropolymer to self-assemble into an enormous variety of highly exotic structures with high control over the detailed shape. This will expand the possibilities of current copolymer technology to the full potential of biopolymers, providing a way of going beyond the current protein- and DNA-based materials~\cite{seeman_dna_review_2003,luo_hydrogel,Gonen2015}.


\begin{thebibliography}{29}%
\makeatletter
\providecommand \@ifxundefined [1]{%
 \@ifx{#1\undefined}
}%
\providecommand \@ifnum [1]{%
 \ifnum #1\expandafter \@firstoftwo
 \else \expandafter \@secondoftwo
 \fi
}%
\providecommand \@ifx [1]{%
 \ifx #1\expandafter \@firstoftwo
 \else \expandafter \@secondoftwo
 \fi
}%
\providecommand \natexlab [1]{#1}%
\providecommand \enquote  [1]{``#1''}%
\providecommand \bibnamefont  [1]{#1}%
\providecommand \bibfnamefont [1]{#1}%
\providecommand \citenamefont [1]{#1}%
\providecommand \href@noop [0]{\@secondoftwo}%
\providecommand \href [0]{\begingroup \@sanitize@url \@href}%
\providecommand \@href[1]{\@@startlink{#1}\@@href}%
\providecommand \@@href[1]{\endgroup#1\@@endlink}%
\providecommand \@sanitize@url [0]{\catcode `\\12\catcode `\$12\catcode
  `\&12\catcode `\#12\catcode `\^12\catcode `\_12\catcode `\%12\relax}%
\providecommand \@@startlink[1]{}%
\providecommand \@@endlink[0]{}%
\providecommand \url  [0]{\begingroup\@sanitize@url \@url }%
\providecommand \@url [1]{\endgroup\@href {#1}{\urlprefix }}%
\providecommand \urlprefix  [0]{URL }%
\providecommand \Eprint [0]{\href }%
\providecommand \doibase [0]{http://dx.doi.org/}%
\providecommand \selectlanguage [0]{\@gobble}%
\providecommand \bibinfo  [0]{\@secondoftwo}%
\providecommand \bibfield  [0]{\@secondoftwo}%
\providecommand \translation [1]{[#1]}%
\providecommand \BibitemOpen [0]{}%
\providecommand \bibitemStop [0]{}%
\providecommand \bibitemNoStop [0]{.\EOS\space}%
\providecommand \EOS [0]{\spacefactor3000\relax}%
\providecommand \BibitemShut  [1]{\csname bibitem#1\endcsname}%
\let\auto@bib@innerbib\@empty
\bibitem [{\citenamefont {Umena}\ \emph {et~al.}(2011)\citenamefont {Umena},
  \citenamefont {Kawakami}, \citenamefont {Shen},\ and\ \citenamefont
  {Kamiya}}]{Umena2011}%
  \BibitemOpen
  \bibfield  {author} {\bibinfo {author} {\bibfnamefont {Y.}~\bibnamefont
  {Umena}}, \bibinfo {author} {\bibfnamefont {K.}~\bibnamefont {Kawakami}},
  \bibinfo {author} {\bibfnamefont {J.-R.}\ \bibnamefont {Shen}}, \ and\
  \bibinfo {author} {\bibfnamefont {N.}~\bibnamefont {Kamiya}},\ }\href@noop {}
  {\bibfield  {journal} {\bibinfo  {journal} {Nature}\ }\textbf {\bibinfo
  {volume} {473}},\ \bibinfo {pages} {55} (\bibinfo {year} {2011})}\BibitemShut
  {NoStop}%
\bibitem [{\citenamefont {Furumi}(2013)}]{furumi::jmatchemc::1::2013}%
  \BibitemOpen
  \bibfield  {author} {\bibinfo {author} {\bibfnamefont {S.}~\bibnamefont
  {Furumi}},\ }\href@noop {} {\bibfield  {journal} {\bibinfo  {journal}
  {Journal of Materials Chemistry C}\ }\textbf {\bibinfo {volume} {1}},\
  \bibinfo {pages} {6003} (\bibinfo {year} {2013})}\BibitemShut {NoStop}%
\bibitem [{\citenamefont {Yang}\ \emph
  {et~al.}(2008{\natexlab{a}})\citenamefont {Yang}, \citenamefont {Bolikal},
  \citenamefont {Becker}, \citenamefont {Kohn}, \citenamefont {Zeiger},\ and\
  \citenamefont {Simon~Jr.}}]{yang::advmat::20::2008}%
  \BibitemOpen
  \bibfield  {author} {\bibinfo {author} {\bibfnamefont {Y.}~\bibnamefont
  {Yang}}, \bibinfo {author} {\bibfnamefont {D.}~\bibnamefont {Bolikal}},
  \bibinfo {author} {\bibfnamefont {M.~L.}\ \bibnamefont {Becker}}, \bibinfo
  {author} {\bibfnamefont {J.}~\bibnamefont {Kohn}}, \bibinfo {author}
  {\bibfnamefont {D.~N.}\ \bibnamefont {Zeiger}}, \ and\ \bibinfo {author}
  {\bibfnamefont {C.~G.}\ \bibnamefont {Simon~Jr.}},\ }\href@noop {} {\bibfield
   {journal} {\bibinfo  {journal} {Advanced Materials}\ }\textbf {\bibinfo
  {volume} {20}},\ \bibinfo {pages} {2037} (\bibinfo {year}
  {2008}{\natexlab{a}})}\BibitemShut {NoStop}%
\bibitem [{\citenamefont {Lodge}(2003)}]{Lodge2003}%
  \BibitemOpen
  \bibfield  {author} {\bibinfo {author} {\bibfnamefont {T.~P.}\ \bibnamefont
  {Lodge}},\ }\href {\doibase 10.1002/macp.200290073} {\bibfield  {journal}
  {\bibinfo  {journal} {Macromolecular Chemistry and Physics}\ }\textbf
  {\bibinfo {volume} {204}},\ \bibinfo {pages} {265} (\bibinfo {year}
  {2003})}\BibitemShut {NoStop}%
\bibitem [{\citenamefont {Capone}\ \emph {et~al.}(2010)\citenamefont {Capone},
  \citenamefont {Hansen},\ and\ \citenamefont {Coluzza}}]{Capone2010}%
  \BibitemOpen
  \bibfield  {author} {\bibinfo {author} {\bibfnamefont {B.}~\bibnamefont
  {Capone}}, \bibinfo {author} {\bibfnamefont {J.-P.}\ \bibnamefont {Hansen}},
  \ and\ \bibinfo {author} {\bibfnamefont {I.}~\bibnamefont {Coluzza}},\ }\href
  {\doibase 10.1039/c0sm00738b} {\bibfield  {journal} {\bibinfo  {journal}
  {Soft Matter}\ }\textbf {\bibinfo {volume} {6}},\ \bibinfo {pages} {6075}
  (\bibinfo {year} {2010})}\BibitemShut {NoStop}%
\bibitem [{\citenamefont {Rothemund}(2006)}]{Rothemund2006}%
  \BibitemOpen
  \bibfield  {author} {\bibinfo {author} {\bibfnamefont {P.~W.~K.}\
  \bibnamefont {Rothemund}},\ }\href {\doibase 10.1038/nature04586} {\bibfield
  {journal} {\bibinfo  {journal} {Nature}\ }\textbf {\bibinfo {volume} {440}},\
  \bibinfo {pages} {297} (\bibinfo {year} {2006})}\BibitemShut {NoStop}%
\bibitem [{\citenamefont {Chino}\ \emph {et~al.}(2015)\citenamefont {Chino},
  \citenamefont {Maglio}, \citenamefont {Nastri}, \citenamefont {Pavone},
  \citenamefont {DeGrado},\ and\ \citenamefont {Lombardi}}]{Chino2015}%
  \BibitemOpen
  \bibfield  {author} {\bibinfo {author} {\bibfnamefont {M.}~\bibnamefont
  {Chino}}, \bibinfo {author} {\bibfnamefont {O.}~\bibnamefont {Maglio}},
  \bibinfo {author} {\bibfnamefont {F.}~\bibnamefont {Nastri}}, \bibinfo
  {author} {\bibfnamefont {V.}~\bibnamefont {Pavone}}, \bibinfo {author}
  {\bibfnamefont {W.~F.}\ \bibnamefont {DeGrado}}, \ and\ \bibinfo {author}
  {\bibfnamefont {A.}~\bibnamefont {Lombardi}},\ }\href {\doibase
  10.1002/ejic.201500470} {\bibfield  {journal} {\bibinfo  {journal} {European
  Journal of Inorganic Chemistry}\ }\textbf {\bibinfo {volume} {2015}},\
  \bibinfo {pages} {3371} (\bibinfo {year} {2015})}\BibitemShut {NoStop}%
\bibitem [{\citenamefont {Huang}\ \emph {et~al.}(2014)\citenamefont {Huang},
  \citenamefont {Oberdorfer}, \citenamefont {Xu}, \citenamefont {Pei},
  \citenamefont {Nannenga}, \citenamefont {Rogers}, \citenamefont {DiMaio},
  \citenamefont {Gonen}, \citenamefont {Luisi},\ and\ \citenamefont
  {Baker}}]{Huang2014a}%
  \BibitemOpen
  \bibfield  {author} {\bibinfo {author} {\bibfnamefont {P.-S.}\ \bibnamefont
  {Huang}}, \bibinfo {author} {\bibfnamefont {G.}~\bibnamefont {Oberdorfer}},
  \bibinfo {author} {\bibfnamefont {C.}~\bibnamefont {Xu}}, \bibinfo {author}
  {\bibfnamefont {X.~Y.}\ \bibnamefont {Pei}}, \bibinfo {author} {\bibfnamefont
  {B.~L.}\ \bibnamefont {Nannenga}}, \bibinfo {author} {\bibfnamefont {J.~M.}\
  \bibnamefont {Rogers}}, \bibinfo {author} {\bibfnamefont {F.}~\bibnamefont
  {DiMaio}}, \bibinfo {author} {\bibfnamefont {T.}~\bibnamefont {Gonen}},
  \bibinfo {author} {\bibfnamefont {B.}~\bibnamefont {Luisi}}, \ and\ \bibinfo
  {author} {\bibfnamefont {D.}~\bibnamefont {Baker}},\ }\href {\doibase
  10.1126/science.1257481} {\bibfield  {journal} {\bibinfo  {journal}
  {Science}\ }\textbf {\bibinfo {volume} {346}},\ \bibinfo {pages} {481}
  (\bibinfo {year} {2014})}\BibitemShut {NoStop}%
\bibitem [{\citenamefont {Conde}\ \emph {et~al.}(2015)\citenamefont {Conde},
  \citenamefont {Oliva}, \citenamefont {Atilano}, \citenamefont {Song},\ and\
  \citenamefont {Artzi}}]{Conde2015}%
  \BibitemOpen
  \bibfield  {author} {\bibinfo {author} {\bibfnamefont {J.}~\bibnamefont
  {Conde}}, \bibinfo {author} {\bibfnamefont {N.}~\bibnamefont {Oliva}},
  \bibinfo {author} {\bibfnamefont {M.}~\bibnamefont {Atilano}}, \bibinfo
  {author} {\bibfnamefont {H.~S.}\ \bibnamefont {Song}}, \ and\ \bibinfo
  {author} {\bibfnamefont {N.}~\bibnamefont {Artzi}},\ }\href@noop {}
  {\bibfield  {journal} {\bibinfo  {journal} {Nature materials}\ } (\bibinfo
  {year} {2015})}\BibitemShut {NoStop}%
\bibitem [{\citenamefont {Gutin}\ and\ \citenamefont
  {Shakhnovich}(1993)}]{Gutin1993a}%
  \BibitemOpen
  \bibfield  {author} {\bibinfo {author} {\bibfnamefont {A.~M.}\ \bibnamefont
  {Gutin}}\ and\ \bibinfo {author} {\bibfnamefont {E.}~\bibnamefont
  {Shakhnovich}},\ }\href@noop {} {\bibfield  {journal} {\bibinfo  {journal}
  {The Journal of Chemical Physics}\ }\textbf {\bibinfo {volume} {98}},\
  \bibinfo {pages} {8174} (\bibinfo {year} {1993})}\BibitemShut {NoStop}%
\bibitem [{\citenamefont {Shakhnovich}\ and\ \citenamefont
  {Gutin}(1993)}]{Shakhnovich1993a}%
  \BibitemOpen
  \bibfield  {author} {\bibinfo {author} {\bibfnamefont {E.~I.}\ \bibnamefont
  {Shakhnovich}}\ and\ \bibinfo {author} {\bibfnamefont {A.~M.}\ \bibnamefont
  {Gutin}},\ }\href@noop {} {\bibfield  {journal} {\bibinfo  {journal}
  {Proceedings of the National Academy of Sciences of the United States of
  America}\ }\textbf {\bibinfo {volume} {90}},\ \bibinfo {pages} {7195}
  (\bibinfo {year} {1993})}\BibitemShut {NoStop}%
\bibitem [{\citenamefont {Pande}\ \emph {et~al.}(2000)\citenamefont {Pande},
  \citenamefont {Grosberg},\ and\ \citenamefont {Tanaka}}]{Pande2000}%
  \BibitemOpen
  \bibfield  {author} {\bibinfo {author} {\bibfnamefont {V.~S.}\ \bibnamefont
  {Pande}}, \bibinfo {author} {\bibfnamefont {A.~Y.}\ \bibnamefont {Grosberg}},
  \ and\ \bibinfo {author} {\bibfnamefont {T.}~\bibnamefont {Tanaka}},\ }\href
  {\doibase http://dx.doi.org/10.1103/RevModPhys.72.259} {\bibfield  {journal}
  {\bibinfo  {journal} {Reviews of Modern Physics}\ }\textbf {\bibinfo {volume}
  {72}},\ \bibinfo {pages} {259} (\bibinfo {year} {2000})}\BibitemShut
  {NoStop}%
\bibitem [{\citenamefont {Gonen}\ \emph {et~al.}(2015)\citenamefont {Gonen},
  \citenamefont {DiMaio}, \citenamefont {Gonen},\ and\ \citenamefont
  {Baker}}]{Gonen2015}%
  \BibitemOpen
  \bibfield  {author} {\bibinfo {author} {\bibfnamefont {S.}~\bibnamefont
  {Gonen}}, \bibinfo {author} {\bibfnamefont {F.}~\bibnamefont {DiMaio}},
  \bibinfo {author} {\bibfnamefont {T.}~\bibnamefont {Gonen}}, \ and\ \bibinfo
  {author} {\bibfnamefont {D.}~\bibnamefont {Baker}},\ }\href@noop {}
  {\bibfield  {journal} {\bibinfo  {journal} {Science}\ }\textbf {\bibinfo
  {volume} {348}},\ \bibinfo {pages} {1365} (\bibinfo {year}
  {2015})}\BibitemShut {NoStop}%
\bibitem [{\citenamefont {Coluzza}(2014)}]{Coluzza2014}%
  \BibitemOpen
  \bibfield  {author} {\bibinfo {author} {\bibfnamefont {I.}~\bibnamefont
  {Coluzza}},\ }\href {\doibase 10.1371/journal.pone.0112852} {\bibfield
  {journal} {\bibinfo  {journal} {PloS one}\ }\textbf {\bibinfo {volume} {9}},\
  \bibinfo {pages} {e112852} (\bibinfo {year} {2014})}\BibitemShut {NoStop}%
\bibitem [{\citenamefont {Khalatur}\ \emph {et~al.}(2007)\citenamefont
  {Khalatur}, \citenamefont {Khokhlov},\ and\ \citenamefont
  {Krotova}}]{Khalatur2007}%
  \BibitemOpen
  \bibfield  {author} {\bibinfo {author} {\bibfnamefont {P.~G.}\ \bibnamefont
  {Khalatur}}, \bibinfo {author} {\bibfnamefont {A.~R.}\ \bibnamefont
  {Khokhlov}}, \ and\ \bibinfo {author} {\bibfnamefont {M.~K.}\ \bibnamefont
  {Krotova}},\ }\href {\doibase 10.1002/masy.200750604} {\bibfield  {journal}
  {\bibinfo  {journal} {Macromolecular Symposia}\ }\textbf {\bibinfo {volume}
  {252}},\ \bibinfo {pages} {36} (\bibinfo {year} {2007})}\BibitemShut
  {NoStop}%
\bibitem [{\citenamefont {Moreno}\ \emph {et~al.}(2013)\citenamefont {Moreno},
  \citenamefont {Lo~Verso}, \citenamefont {Sanchez-Sanchez}, \citenamefont
  {Arbe}, \citenamefont {Colmenero},\ and\ \citenamefont
  {Pomposo}}]{moreno2013advantages}%
  \BibitemOpen
  \bibfield  {author} {\bibinfo {author} {\bibfnamefont {A.~J.}\ \bibnamefont
  {Moreno}}, \bibinfo {author} {\bibfnamefont {F.}~\bibnamefont {Lo~Verso}},
  \bibinfo {author} {\bibfnamefont {A.}~\bibnamefont {Sanchez-Sanchez}},
  \bibinfo {author} {\bibfnamefont {A.}~\bibnamefont {Arbe}}, \bibinfo {author}
  {\bibfnamefont {J.}~\bibnamefont {Colmenero}}, \ and\ \bibinfo {author}
  {\bibfnamefont {J.~A.}\ \bibnamefont {Pomposo}},\ }\href@noop {} {\bibfield
  {journal} {\bibinfo  {journal} {Macromolecules}\ }\textbf {\bibinfo {volume}
  {46}},\ \bibinfo {pages} {9748} (\bibinfo {year} {2013})}\BibitemShut
  {NoStop}%
\bibitem [{\citenamefont {Coluzza}\ \emph
  {et~al.}(2013{\natexlab{a}})\citenamefont {Coluzza}, \citenamefont {van
  Oostrum}, \citenamefont {Capone}, \citenamefont {Reimhult},\ and\
  \citenamefont {Dellago}}]{Coluzza2013}%
  \BibitemOpen
  \bibfield  {author} {\bibinfo {author} {\bibfnamefont {I.}~\bibnamefont
  {Coluzza}}, \bibinfo {author} {\bibfnamefont {P.}~\bibnamefont {van
  Oostrum}}, \bibinfo {author} {\bibfnamefont {B.}~\bibnamefont {Capone}},
  \bibinfo {author} {\bibfnamefont {E.}~\bibnamefont {Reimhult}}, \ and\
  \bibinfo {author} {\bibfnamefont {C.}~\bibnamefont {Dellago}},\ }\href
  {\doibase 10.1103/PhysRevLett.110.075501} {\bibfield  {journal} {\bibinfo
  {journal} {Physical Review Letters}\ }\textbf {\bibinfo {volume} {110}},\
  \bibinfo {pages} {075501} (\bibinfo {year} {2013}{\natexlab{a}})}\BibitemShut
  {NoStop}%
\bibitem [{\citenamefont {Coluzza}\ \emph
  {et~al.}(2013{\natexlab{b}})\citenamefont {Coluzza}, \citenamefont {van
  Oostrum}, \citenamefont {Capone}, \citenamefont {Reimhult},\ and\
  \citenamefont {Dellago}}]{Coluzza2012c}%
  \BibitemOpen
  \bibfield  {author} {\bibinfo {author} {\bibfnamefont {I.}~\bibnamefont
  {Coluzza}}, \bibinfo {author} {\bibfnamefont {P.~D.}\ \bibnamefont {van
  Oostrum}}, \bibinfo {author} {\bibfnamefont {B.}~\bibnamefont {Capone}},
  \bibinfo {author} {\bibfnamefont {E.}~\bibnamefont {Reimhult}}, \ and\
  \bibinfo {author} {\bibfnamefont {C.}~\bibnamefont {Dellago}},\ }\href@noop
  {} {\bibfield  {journal} {\bibinfo  {journal} {Soft Matter}\ }\textbf
  {\bibinfo {volume} {9}},\ \bibinfo {pages} {938} (\bibinfo {year}
  {2013}{\natexlab{b}})}\BibitemShut {NoStop}%
\bibitem [{\citenamefont {Wang}\ \emph
  {et~al.}(2008{\natexlab{a}})\citenamefont {Wang}, \citenamefont {Luo},
  \citenamefont {Liu},\ and\ \citenamefont {Huang}}]{ISI:000253221300007}%
  \BibitemOpen
  \bibfield  {author} {\bibinfo {author} {\bibfnamefont {G.}~\bibnamefont
  {Wang}}, \bibinfo {author} {\bibfnamefont {X.}~\bibnamefont {Luo}}, \bibinfo
  {author} {\bibfnamefont {C.}~\bibnamefont {Liu}}, \ and\ \bibinfo {author}
  {\bibfnamefont {J.}~\bibnamefont {Huang}},\ }\href {\doibase
  10.1002/pola.22550} {\bibfield  {journal} {\bibinfo  {journal} {Journal of
  Polymer Science Part A: Polymer Chemistry}\ }\textbf {\bibinfo {volume}
  {46}},\ \bibinfo {pages} {2154} (\bibinfo {year}
  {2008}{\natexlab{a}})}\BibitemShut {NoStop}%
\bibitem [{\citenamefont {Wang}\ \emph
  {et~al.}(2008{\natexlab{b}})\citenamefont {Wang}, \citenamefont {Luo},
  \citenamefont {Liu},\ and\ \citenamefont {Huang}}]{ISI:000254135200022}%
  \BibitemOpen
  \bibfield  {author} {\bibinfo {author} {\bibfnamefont {G.}~\bibnamefont
  {Wang}}, \bibinfo {author} {\bibfnamefont {X.}~\bibnamefont {Luo}}, \bibinfo
  {author} {\bibfnamefont {C.}~\bibnamefont {Liu}}, \ and\ \bibinfo {author}
  {\bibfnamefont {J.}~\bibnamefont {Huang}},\ }\href {\doibase
  10.1002/pola.22550} {\bibfield  {journal} {\bibinfo  {journal} {Journal of
  Polymer Science Part A: Polymer Chemistry}\ }\textbf {\bibinfo {volume}
  {46}},\ \bibinfo {pages} {2154} (\bibinfo {year}
  {2008}{\natexlab{b}})}\BibitemShut {NoStop}%
\bibitem [{\citenamefont {Yang}\ \emph
  {et~al.}(2008{\natexlab{b}})\citenamefont {Yang}, \citenamefont {Zhou},
  \citenamefont {Shi}, \citenamefont {Wang},\ and\ \citenamefont
  {Pan}}]{ISI:000259589300026}%
  \BibitemOpen
  \bibfield  {author} {\bibinfo {author} {\bibfnamefont {L.}~\bibnamefont
  {Yang}}, \bibinfo {author} {\bibfnamefont {H.}~\bibnamefont {Zhou}}, \bibinfo
  {author} {\bibfnamefont {G.}~\bibnamefont {Shi}}, \bibinfo {author}
  {\bibfnamefont {Y.}~\bibnamefont {Wang}}, \ and\ \bibinfo {author}
  {\bibfnamefont {C.-Y.}\ \bibnamefont {Pan}},\ }\href {\doibase
  10.1002/pola.22975} {\bibfield  {journal} {\bibinfo  {journal} {Journal of
  Polymer Science Part A: Polymer Chemistry}\ }\textbf {\bibinfo {volume}
  {46}},\ \bibinfo {pages} {6641} (\bibinfo {year}
  {2008}{\natexlab{b}})}\BibitemShut {NoStop}%
\bibitem [{\citenamefont {Seeman}(2003)}]{seeman_dna_review_2003}%
  \BibitemOpen
  \bibfield  {author} {\bibinfo {author} {\bibfnamefont {N.~C.}\ \bibnamefont
  {Seeman}},\ }\href@noop {} {\bibfield  {journal} {\bibinfo  {journal}
  {Nature}\ }\textbf {\bibinfo {volume} {421}},\ \bibinfo {pages} {427}
  (\bibinfo {year} {2003})}\BibitemShut {NoStop}%
\bibitem [{\citenamefont {Lee}\ \emph {et~al.}(2012)\citenamefont {Lee},
  \citenamefont {Peng}, \citenamefont {Yang}, \citenamefont {Roh},
  \citenamefont {Funabashi}, \citenamefont {Park}, \citenamefont {Rice},
  \citenamefont {Chen}, \citenamefont {Long}, \citenamefont {Wu},\ and\
  \citenamefont {Luo}}]{luo_hydrogel}%
  \BibitemOpen
  \bibfield  {author} {\bibinfo {author} {\bibfnamefont {J.~B.}\ \bibnamefont
  {Lee}}, \bibinfo {author} {\bibfnamefont {S.}~\bibnamefont {Peng}}, \bibinfo
  {author} {\bibfnamefont {D.}~\bibnamefont {Yang}}, \bibinfo {author}
  {\bibfnamefont {Y.~H.}\ \bibnamefont {Roh}}, \bibinfo {author} {\bibfnamefont
  {H.}~\bibnamefont {Funabashi}}, \bibinfo {author} {\bibfnamefont
  {N.}~\bibnamefont {Park}}, \bibinfo {author} {\bibfnamefont {E.~J.}\
  \bibnamefont {Rice}}, \bibinfo {author} {\bibfnamefont {L.}~\bibnamefont
  {Chen}}, \bibinfo {author} {\bibfnamefont {R.}~\bibnamefont {Long}}, \bibinfo
  {author} {\bibfnamefont {M.}~\bibnamefont {Wu}}, \ and\ \bibinfo {author}
  {\bibfnamefont {D.}~\bibnamefont {Luo}},\ }\href@noop {} {\bibfield
  {journal} {\bibinfo  {journal} {Nature Nanotechnology}\ }\textbf {\bibinfo
  {volume} {7}},\ \bibinfo {pages} {816} (\bibinfo {year} {2012})}\BibitemShut
  {NoStop}%
\bibitem [{\citenamefont {Miyazawa}\ and\ \citenamefont
  {Jernigan}(1996)}]{Miyazawa1996}%
  \BibitemOpen
  \bibfield  {author} {\bibinfo {author} {\bibfnamefont {S.}~\bibnamefont
  {Miyazawa}}\ and\ \bibinfo {author} {\bibfnamefont {R.~L.}\ \bibnamefont
  {Jernigan}},\ }\href@noop {} {\bibfield  {journal} {\bibinfo  {journal}
  {Journal of Molecular Biology}\ }\textbf {\bibinfo {volume} {256}},\ \bibinfo
  {pages} {623} (\bibinfo {year} {1996})}\BibitemShut {NoStop}%
\bibitem [{\citenamefont {Betancourt}\ and\ \citenamefont
  {Thirumalai}(1999)}]{Betancourt1999}%
  \BibitemOpen
  \bibfield  {author} {\bibinfo {author} {\bibfnamefont {M.~R.}\ \bibnamefont
  {Betancourt}}\ and\ \bibinfo {author} {\bibfnamefont {D.}~\bibnamefont
  {Thirumalai}},\ }\href {\doibase 10.1110/ps.8.2.361} {\bibfield  {journal}
  {\bibinfo  {journal} {Protein Science}\ }\textbf {\bibinfo {volume} {8}},\
  \bibinfo {pages} {361} (\bibinfo {year} {1999})}\BibitemShut {NoStop}%
\bibitem [{\citenamefont {Coluzza}(2011)}]{Coluzza2011}%
  \BibitemOpen
  \bibfield  {author} {\bibinfo {author} {\bibfnamefont {I.}~\bibnamefont
  {Coluzza}},\ }\href {\doibase 10.1371/journal.pone.0020853} {\bibfield
  {journal} {\bibinfo  {journal} {PloS one}\ }\textbf {\bibinfo {volume} {6}},\
  \bibinfo {pages} {e20853} (\bibinfo {year} {2011})}\BibitemShut {NoStop}%
\bibitem [{\citenamefont {Irb{\"{a}}ck}\ \emph {et~al.}(2000)\citenamefont
  {Irb{\"{a}}ck}, \citenamefont {Sjunnesson},\ and\ \citenamefont
  {Wallin}}]{Irback2000}%
  \BibitemOpen
  \bibfield  {author} {\bibinfo {author} {\bibfnamefont {A.}~\bibnamefont
  {Irb{\"{a}}ck}}, \bibinfo {author} {\bibfnamefont {F.}~\bibnamefont
  {Sjunnesson}}, \ and\ \bibinfo {author} {\bibfnamefont {S.}~\bibnamefont
  {Wallin}},\ }\href {\doibase 10.1073/pnas.240245297} {\bibfield  {journal}
  {\bibinfo  {journal} {Proceedings of the National Academy of Sciences of the
  United States of America}\ }\textbf {\bibinfo {volume} {97}},\ \bibinfo
  {pages} {13614} (\bibinfo {year} {2000})}\BibitemShut {NoStop}%
\bibitem [{\citenamefont {Coluzza}\ and\ \citenamefont
  {Frenkel}(2005)}]{Coluzza2005}%
  \BibitemOpen
  \bibfield  {author} {\bibinfo {author} {\bibfnamefont {I.}~\bibnamefont
  {Coluzza}}\ and\ \bibinfo {author} {\bibfnamefont {D.}~\bibnamefont
  {Frenkel}},\ }\href {\doibase 10.1002/cphc.200400629} {\bibfield  {journal}
  {\bibinfo  {journal} {Chemphyschem : a European journal of chemical physics
  and physical chemistry}\ }\textbf {\bibinfo {volume} {6}},\ \bibinfo {pages}
  {1779} (\bibinfo {year} {2005})}\BibitemShut {NoStop}%
\bibitem [{\citenamefont {Finkelstein}\ and\ \citenamefont
  {Ptitsyn}(2002)}]{finkelstein2002protein}%
  \BibitemOpen
  \bibfield  {author} {\bibinfo {author} {\bibfnamefont {A.~V.}\ \bibnamefont
  {Finkelstein}}\ and\ \bibinfo {author} {\bibfnamefont {O.}~\bibnamefont
  {Ptitsyn}},\ }\href@noop {} {\emph {\bibinfo {title} {Protein physics: a
  course of lectures}}}\ (\bibinfo  {publisher} {Academic Press},\ \bibinfo
  {year} {2002})\BibitemShut {NoStop}%
\end{thebibliography}
%

\clearpage
\section{Acknowledgment}
We would like to thank Achille Giacometti and Tatjana \v{S}krbi\'c for fruitful discussions. We acknowledge support from the Austrian Science Fund (FWF) project P23846-N16, the Mahlke-Obermann Stiftung and the European Union's Seventh Framework Programme for research, technological development and demonstration under grant agreement no 609431,  VSC Research Center funded by the Austrian Federal Ministry of Science, Research and Economy (bmwfw). The computational results presented have been achieved using the Vienna Scientific Cluster (VSC).

\section{Methods}\label{sec:methods}

\begin{figure*}
\leftskip 1.7 cm 
 {\bf (a)\hspace{7.5 cm}(b)}\\
 \centering
\includegraphics[width=0.85\textwidth]{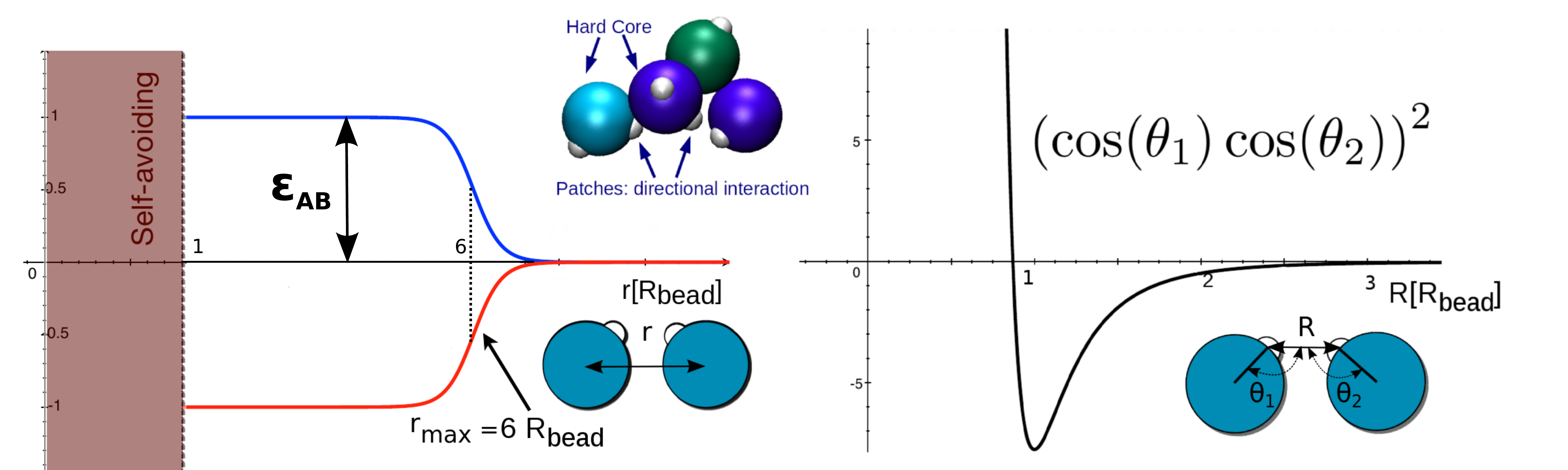} \\
\includegraphics[width=0.75\textwidth]{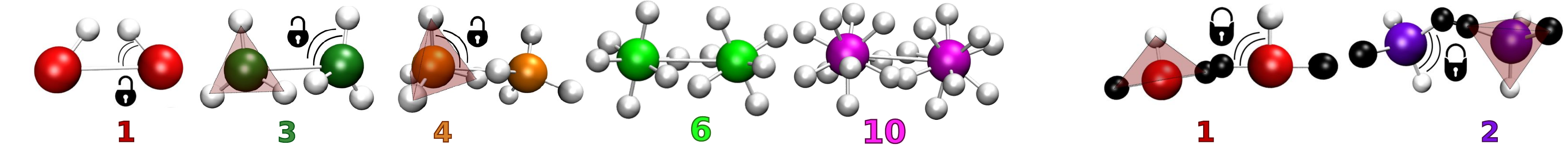} \\
\caption{\label{fig:potentials}Interaction potentials. a) Isotropic square-well like interaction as a function of the distance of the centres of the spherical monomers $r$ (see left inset). The well depth is controlled by a pre-factor $\epsilon_{AB}$, which is different for each different pair of monomers. The pre-factors are grouped in an interaction matrix, the size of which depends on the number of monomer types in the range 3 to 20 (find the matrices used in the Supplementary Informations). For an alphabet of size $N$, the set of $N(N+1)/2$ heterogeneous interactions are Gaussian random numbers with zero average and standard deviation of $0.33 k_BT_\textrm{{Ref}}$, similarly to protein coarse-grained protein models~\cite{Coluzza2014,Miyazawa1996,Betancourt1999}, where $T_\textrm{{Ref}}$ is a reference temperature that sets the scale of interactions. Inset: schematic illustration of patchy polymers. Small white spheres indicate the patches. b) Directional potential. The radial Lennard-Jones contribution in the plot is multiplied by the directional contribution $(\cos{\theta_1}\cos{\theta_2})^2$, where $\theta_1$ and $\theta_2$  are the angles between the patch vector and the distance vector (right inset).}
\end{figure*} 
A guiding theory for heteropolymer designability is given by the Random Energy Model (REM)~\cite{Gutin1993a,Shakhnovich1993a,Pande2000}.
A clear review can be found in the seminal works of Pande \textit{et al.}~\cite{Pande2000}, where it is shown how the designability of a heteropolymer increases with the total number of possible bonds for each bead (valence) and decreases with the conformational entropy per bead. Hence, it is reasonable to assume that directionality (the patches) combined with isotropic interactions would increase designability, because the valence (i.e. the total number of possible bonds) remains constant, while the conformational entropy per bead decreases. In fact, the introduction of the patches decreases the entropy by favouring the system to populate more specific structures with the patches along particular directions. 

On the other hand, if the number of patches increases to much, the interactions become again close to isotropic and the designability decreases again.
Hence, a model is needed that explicitly brings about the designability from a basic heteropolymer model by controlling the alphabet size and the conformational entropy per particle. 

Here we will use the patchy polymer model~\cite{Coluzza2013,Coluzza2012c} that has already been proven to be effective to refold artificial sequences into unique target structures for some specific cases of configurations of the patches. With patchy polymers we are able to change  both the alphabet size of the isotropic interactions, the number of directional interactions incrementing them from the heteropolymer limit without patches, and finally change also the geometrical arrangement of the patches. The isotropic interaction energy $E_{AB}\left(r\right)$ between two different sub-units of types A and B is represented  as a simple-square-well like shape (Fig.~\ref{fig:potentials})
\begin{equation} 
E_{AB}\left(r\right)= \begin{cases} 
\epsilon_{AB}\left[1-\dfrac{1}{1.0 + \textrm{e}^{2.5 \left(r_{max} - r\right)}}\right] &\mbox{if } r > R_{bead} \\ 
\infty, =&\mbox{if } r < R_{bead} 
\end{cases}\label{CalphaInt}
\end{equation}

where $r$ is the distance of the centres of the beads and $R_{bead}$ is the hard core radius, which is the same for each bead. $\epsilon_{AB}$ is a different pre-factor for every different pair of monomers. The cut-off distance $r_{max}=6 R_{bead}$ is the distance at which $E_{AB}=\epsilon_{AB}/2$ and was derived with a trial and error approach on coarse-grained proteins in the Caterpillar protein model~\cite{Coluzza2014,Coluzza2011}. 

As directional interaction between the patches we employ the potential derived by Irb\"{a}ck et al.~\cite{Irback2000}, commonly used to model hydrogen bonds. It is represented by a 10-12 Lennard-Jones type potential multiplied by a factor containing the angles between the patches and the bead radius (Fig.~\ref{fig:potentials}), so that the energy is minimum if the patches face each others (when they are opposite to each others the radial part of the potential is $\sim 0$)
\begin{equation}
E_{p}=s\,\epsilon_{p}
\left(\cos{\theta_1}\cos{\theta_2}\right)^\nu
\left[5\left(\frac{\sigma}{R}\right)^{12}-
      6\left(\frac{\sigma}{R}\right)^{10}\right].
\label{Conformationalenergy}
\end{equation}
Here $R$ is the distance of the patches as in Fig.~\ref{fig:potentials} (right inset), $\epsilon_{p}=3.1~k_BT$ and $\nu=2$~\cite{Irback2000} while we set $\sigma=R_{bead}$.
The scaling factor $s$ is chosen to not over favour the patch contribution over the isotropic one. If its value is too large all sequences form regular structures that depend solely on the symmetries of the patch arrangements on the beads. On the other hand, if it is too small all sequences fail to self-assemble and collapse into random glassy three dimensional structures. A good number was found to be $4$ in Ref.~\cite{Coluzza2012c}. The neighbour beads along the chain are bonded via a harmonic spring potential.\\
In order to find if the system is designable or not, we identify for each configuration of the patches and an alphabet size at least one pattern (sequence) that has a global free energy minimum into a given structure. Hence, we have to find such a pattern (via DESIGN MC simulation~\cite{Coluzza2012c}) and verify if it is capable of folding (via the FOLDING MC simulation~\cite{Coluzza2012c}). In all Monte Carlo simulations we enhance the sampling with the Virtual Move Parallel Tempering algorithm~\cite{Coluzza2005}, performing each simulation at 16 different temperatures in the set [3, 2.5, 2.0, 1.6, 1.4, 1.2, 1.0, 0.9, 0.8, 0.75, 0.7, 0.65, 0.6, 0.55, 0.5, 0.4].
To increase the chances to find such a pattern, we first perform a SEEK MC simulation~\cite{Coluzza2012c}, in order to find potentially designable target structures (see Supplementary Figure~\ref{seek}). The target structures are represented for each case in Fig.~\ref{fig:diagram}. Then we perform the DESIGN, where we explore different sequences while keeping the target structure frozen. Here we choose the optimized sequence in the global minimum of the free energy, which corresponds in our method to a low potential energy and a high heterogeneity of the sequence~\cite{Coluzza2012c}. 
Starting from a fully stretched structure we then perform a FOLDING Monte Carlo simulation, in order to study the self-assembling properties of this pattern. Here we explore the conformational space keeping the pattern fixed in the designed sequence. We project the FOLDING free energy onto an order parameter, namely the root mean square displacement of the inter-particle distance ($DRMSD$) between the target structure and each sampled structure:
\begin{equation}
DRMSD = \frac{1}{N} \sqrt{\sum_{ij} (|\Delta \vec{r_{ij}}| - |\Delta \vec{r_{ij}}^T| )^2 }
\end{equation}
where $\vec{r_{ij}}$ is the distance between the sphere $i$ and $j$ while $\vec{r_{ij}}^T$ is the same distance calculated over the target structure, and $N$ is the chain length (50 in our case).
The $DRMSD$ has already been shown to be a proper order parameter to study the folding process~\cite{Coluzza2014}. $DRMSD=0$ corresponds uniquely to the target structure. The closer the global minimum is to $DRMSD=0$, the smaller the corresponding ensemble of structures. Thus, if the free energy landscape has a clear global minimum close to $DRMSD=0$, we can identify at least one pattern that drives the system to fold into a unique target structure: the configuration is designable. However, not all global minima correspond to folded conformations. Indeed, the different cases can be discriminated only by increasing the temperature and pushing the system to unfold. If the chain at low temperature is folded, then there will be an intermediate temperature at which the system explore conformations with higher values of $DRMSD$ corresponding to disordered globular structures (molten globule). The progressive unfolding results in either the appearance of a second minimum (Supplementary Figure~\ref{fig:highT}) or the spread of the width of the initial global minimum.  On the other hand, for patterns that do not have a folded state the equilibrium configuration is always a molten globule and the temperature increase does not significantly change the position of the minimum, which corresponds to values of $DRMSD\simeq1:1.5$ in grey in Fig.~\ref{fig:results}. If the system does not reach the folded state, the SDF trials fail, then associated to the tested configuration of the patches there might be only a handful of structures that are designable, if they exists at all. A heteropolymer with few and hard-to-find designable structures is anyhow not a good candidate for potential applications and is categorized as not-designable. \\
The radial distribution functions of the patchy polymers in Fig.~\ref{fig:gr}a $g(r)$ has been averaged on the 20 most designable structures for every case. For proteins in Fig.~\ref{fig:gr}b the $g(R)$ is calculated for the C$_{\alpha}$ in three characteristics examples out of 20 analysed natural proteins, each averaged on multiple equilibrium PDB (Protein Data Bank). The other analysed proteins are not shown but give similar $g(R)$. The normalization has been performed on the same ideal gas with average density, to make the $g(r)$ of proteins with different length comparable. All the $g(r)$ have been calculated by neglecting the contribution of the beads (or amino acids) directly connected along the chain, in order to ignore their trivial contribution to the first neighbours peak. 
\clearpage
\appendix

\renewcommand\thefigure{\thesection.\arabic{figure}}
\setcounter{figure}{0}

\section*{SUPPLEMENTARY MATERIAL}
\renewcommand\thefigure{S.\arabic{figure}}

We use Monte Carlo simulations to investigate patchy polymers composed by $50$ patchy particles 
decorated with $n$ patches, where $n=1,2$ for freely jointed chains and $n=0,1,3,4,6,10$ for freely rotating chains. The patches are equispaced on the equator and placed on the vertices of a equilateral triangle and a tetrahedron for the $n=3$ and $n=4$ cases, respectively. For $n>4$ the patches are placed on the surface in the most symmetric 
way by using the following numerical procedure:
\begin{enumerate}
\item $n$ patches are randomly placed on a sphere, their positions given by the set of vectors $\{\vec{r_1},\ldots,\vec{r_n}\}$ 
\item we assign a fictitious energy to the system, defined as $U = \frac{1}{2} \sum_{i\neq j} |\vec{r_i} - \vec{r_j}|^{-1}$
\item we minimise $U$ by attempting to move a randomly chosen patch, accepting the move if the total 
energy of the system consequently decreases. Formally, this can be regarded as a Monte Carlo (MC) 
simulation performed at temperature zero.
\item We iterate this procedure until convergence of $U$. 
\end{enumerate}
For completeness, we note that the above method produces a patch distribution which is independent of the 
definition of distance between two patches, being it the Euclidean distance or a spherical distance, for 
all the values of $n$ considered here. In addition, the patch distribution makes sure that two particles
cannot be involved in more than one bond. 
\begin{figure*}[htp]
\includegraphics[width=0.7\columnwidth]{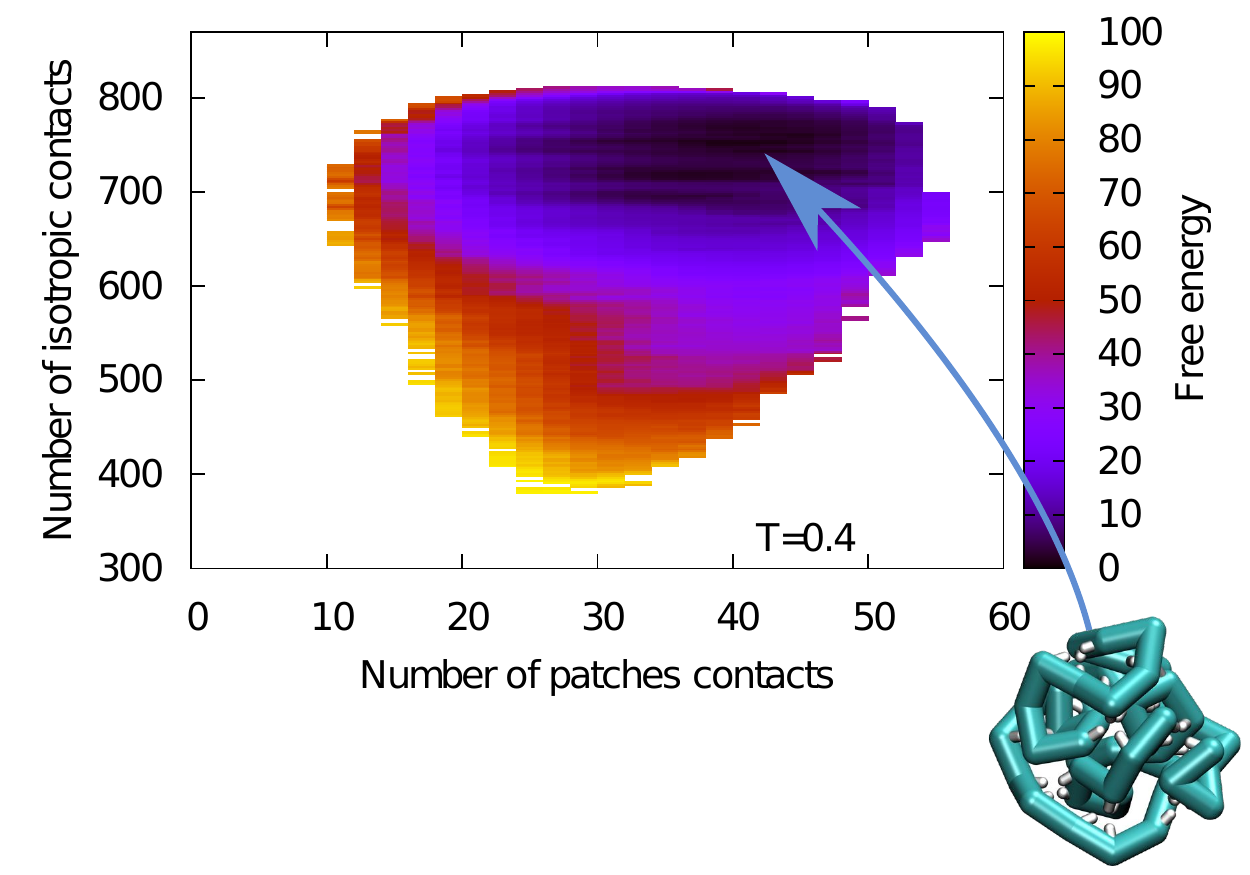}
\caption{Free energy landscape sampled by SEEK for one patch and alphabet 3 in the freely rotating chain model. The free energy is in function of the total number of contacts between the spheres (distance below $6~R_{bead}$) and the total number of contacts between the patches (distance below $1.25~R_{bead}$ and angles $\theta_1$ and $\theta_2$ $> 0.8~\pi$). The target structure is chosen in the global minimum of this landscape. Following the above definition of patches contacts, in the target structure the $80\%$ of the patches are maximally oriented between each others. However, we observe by looking closer at the structure, that all the patches are interacting. Nevertheless, the close packing peak in the radial distribution function dominates on the directional interaction peak. Hence, the close packing is not suppressed even when all the directional interactions are fulfilled. Thus, increasing the relative strength of the directional interactions will not make the first peak disappear. }
\label{seek}
\end{figure*}

\begin{figure*}
\includegraphics[width=0.6\columnwidth]{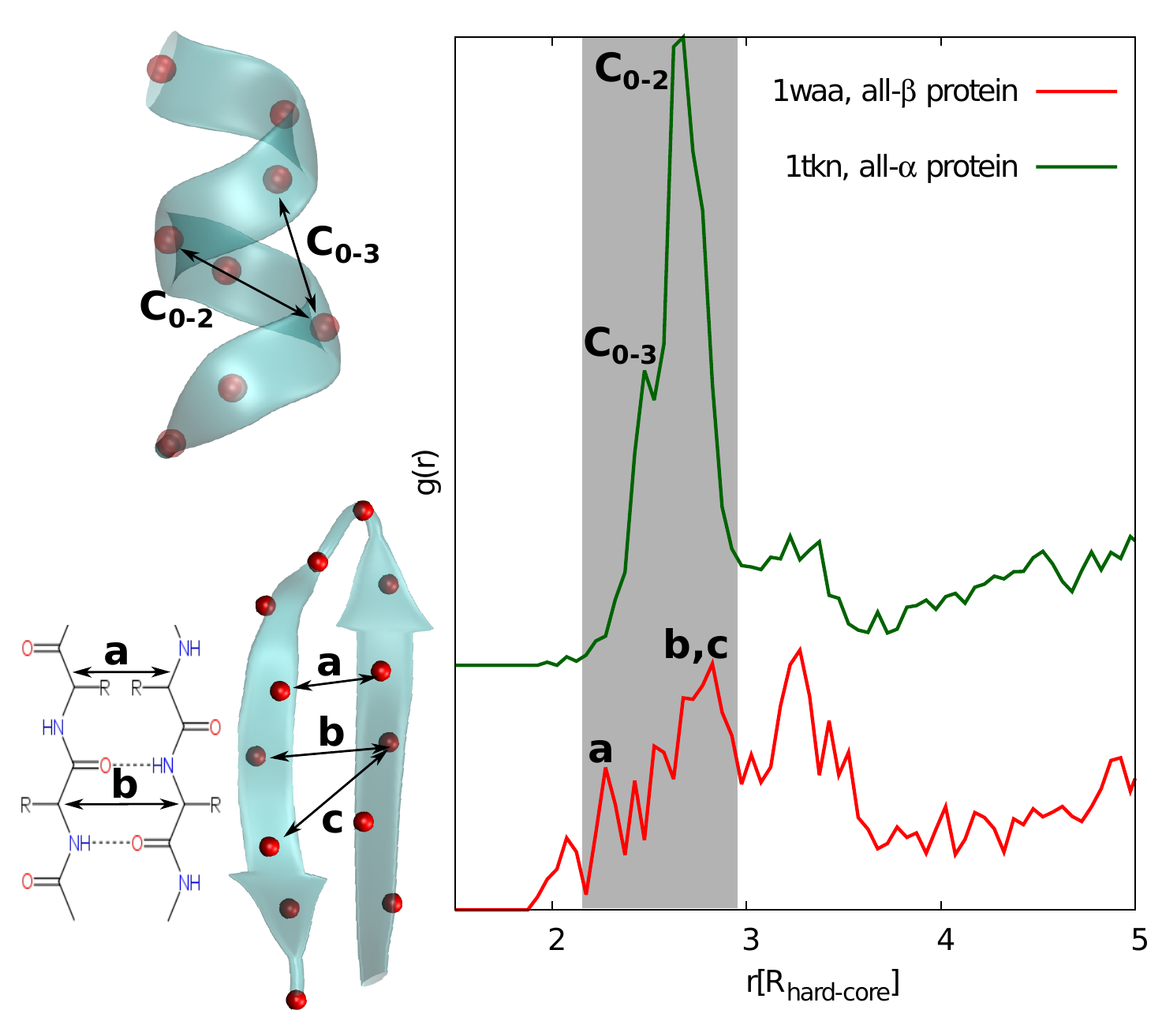}
\caption{ Radial distribution functions ($g(r)$) of proteins. The $g(r)$ are pair distribution functions between the C$_{\alpha}$ in two characteristics examples out of 20 analysed natural proteins, each averaged on multiple equilibrium PDB (Protein Data Bank). The peaks correspond to the typical distances imposed by the secondary structure~\cite{finkelstein2002protein}. In $\alpha$-helix they are 5.0\AA (C$_{0-2}$ in figure), 5.4\AA (C$_{0-3}$ in figure). The peak outside from the grey area corresponds to the fourth neighbour along the chain. In anti-parallel $\beta$-sheets the peaks \textit{a} and \textit{b} correspond to the C$_{\alpha}$ facing each others. The peak outside from the grey area correspond to the second neighbour along the strand. }
\label{fig:protein_scheme}
\end{figure*}

\begin{figure*}
\includegraphics[width=1\columnwidth]{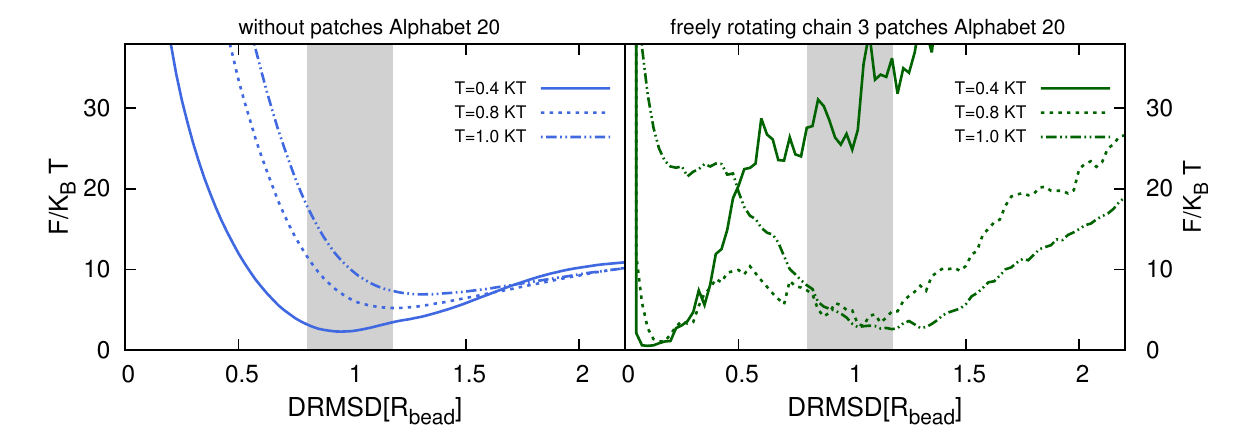}
\caption{FOLDING free energy landscape as a function of the distance root mean square displacement ($DRMSD$) for an example of non-designable system (left) and designable system (right). By increasing the temperature, the position of the minimum of the non-designable system does not change significantly, while in the designable system we observe a significant shift of the minimum due to the disordered globular structures (molten globule). }
\label{fig:highT}
\end{figure*}

\begin{figure*}[h!]
\subfigure[][{}]
{\includegraphics[width=0.45\textwidth]{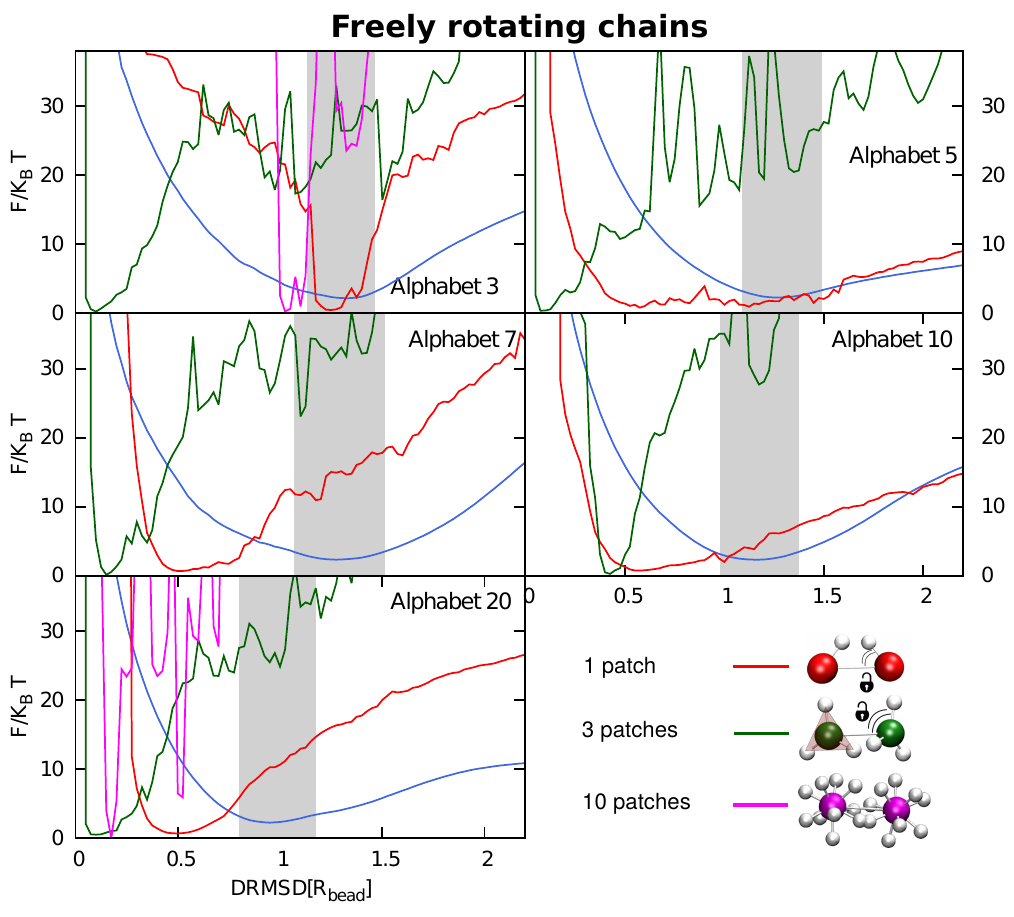}
\label{fig:total_results_free}}
\subfigure[][{}]
{\includegraphics[width=0.465\textwidth]{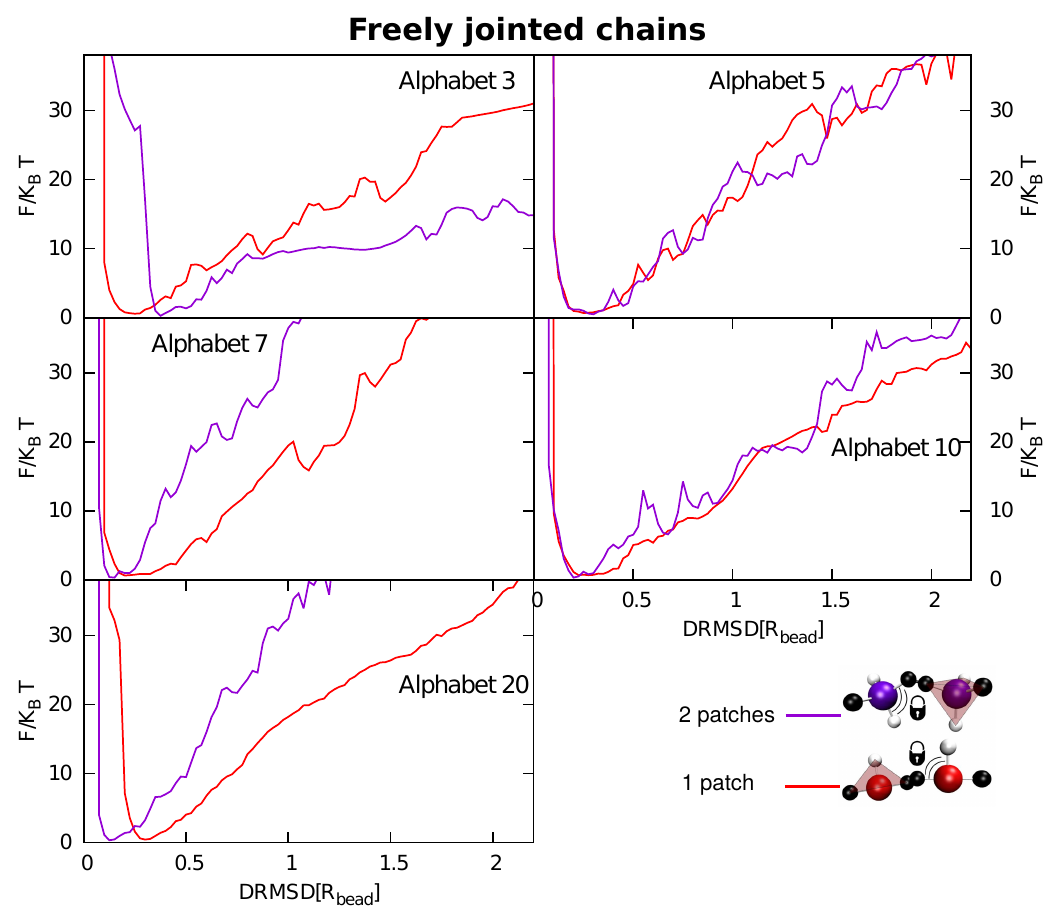}
\label{fig:total_results_const}}
\caption{\small  FOLDING free energy landscapes. The free energy is plotted as a function of the distance root mean square displacement ($DRMSD$) for freely rotating chain (left) and freely jointed chain (right), for different patches numbers and alphabet sizes at temperature $0.4$. }
\label{fig:folding_total_results}
\end{figure*}

\end{document}